\documentclass[conference]{IEEEtran}
\usepackage{graphicx}
\usepackage{subfigure}
\usepackage{amsmath}
\usepackage{amsbsy}
\usepackage{amssymb}
\usepackage{graphicx,xparse}
\usepackage{mathtools}
\usepackage{multirow}
\usepackage{array}
\usepackage{caption}
\usepackage[utf8]{inputenc}
\usepackage{acronym}
\usepackage{mathptmx}
\usepackage{enumitem}
\setlist[itemize]{noitemsep, topsep=0pt}
\setlist[enumerate]{noitemsep, topsep=0pt}
\usepackage{longtable}
\usepackage[hidelinks]{hyperref}
\usepackage{xcolor}
\hypersetup{
    colorlinks,
    linkcolor={black!50!black},
    citecolor={black!90!black},
    urlcolor={black!80!black}
}
\usepackage{xcolor}
\usepackage{cite}
\usepackage{comment}
\usepackage[acronym]{glossaries}

\newacronym{6g}{6G}{sixth generation}
\newacronym{ris}{RIS}{reconfigurable intelligent surfaces}
\newacronym{dnn}{DNN}{Deep Neural Network}
\newacronym{ann}{ANN}{Artificial Neural Network}
\newacronym{so}{SO}{scattering objects}
\newacronym{bilstm}{biLSTM}{bidirectional long-short term memory}
\newacronym{lstm}{LSTM}{long-short term memory}
\newacronym{ue}{UE}{user equipment}
\newacronym{mse}{MSE}{mean squared error}
\newacronym{bs}{BS}{base station}

\title{Reconfigurable Intelligent Surfaces in Dynamic Rich Scattering Environments: BiLSTM-Based Optimization for Accurate User Localization}
\author{Anum~Umer, Ivo~M\"{u}\"{u}rsepp, Muhammad~Mahtab~Alam,~\IEEEmembership{Senior~Member,~IEEE} \\
\IEEEauthorblockA{\textit{Thomas Johann Seebeck Department of Electronics} \\
\textit{Tallinn University of Technology}\\
Tallinn, Estonia \\
\{anum.umer, ivo.muursepp, muhammad.alam\}@taltech.ee}
}

\begin{document}
\maketitle

%


\begin{abstract}
The integration of \acrfull{ris} in wireless environments offers channel programmability and dynamic control over propagation channels, which is expected to play a crucial role in \acrfull{6g} networks. The majority of RIS-related research has focused on simpler, quasi-free-space conditions, where wireless channels are typically modeled analytically. However, many practical localization scenarios unfold in environments characterized by rich scattering that also change over time. These dynamic and complex conditions pose significant challenges in determining the optimal RIS configuration to maximize localization accuracy. In this paper, we present our approach to overcoming this challenge. This paper introduces a novel approach that leverages a \acrfull{bilstm} network, trained with a simulator that accurately reflects wave physics, to capture the relationship between wireless channels and the RIS configuration under dynamic, rich-scattering conditions. We use this approach to optimize RIS configurations for enhanced \acrfull{ue} localization, measured by \acrfull{mse}. Through extensive simulations, we demonstrate that our approach adapts RIS configurations to significantly improve localization accuracy in such dynamically changing rich scattering environments. 
\end{abstract}


\section{Introduction}
The \acrfull{ris} are increasingly recognized for their potential in reshaping wireless communications by converting traditionally unpredictable propagation channels into controllable systems \cite{ref1}. These dynamic metasurfaces, consisting of numerous low-cost scattering elements, can be tuned almost passively to optimize signal propagation. From a localization and sensing perspective, RIS significantly enhance the accuracy of \acrfull{ue} positioning and environment sensing by managing multipath effects and reducing signal degradation \cite{ref2}. This capability improves signal integrity and strengthens the reliability of localization and sensing techniques essential for applications requiring high spatial accuracy, such as automated navigation and context-aware services \cite{ref4}. The dynamic nature of RIS technology allows real-time adaptation to environmental changes, enhancing both communication and localization performance, which is crucial for robust, efficient communication systems in scenarios characterized by rich scattering and dynamic changes typical of emerging network technologies \cite{ref3}. Understanding the optimal configurations of RIS in such scenarios is critical for realizing the full potential of next-generation communication networks, particularly for enhancing precise localization capabilities of \acrfull{6g} networks.

Recent studies on \acrshort{ris} assisted communication and localization networks have predominantly focused on quasi-free-space contexts with minimal known scatterers, which allow for the application of traditional analytical channel models for wireless propagation. Most algorithmic explorations presume that channel conditions are known, but consistently acquiring this information entails substantial overhead \cite{ref6}, leading to innovative approaches to employ RIS without deterministic channel data \cite{ref7}. The challenge of accurately estimating channels within RIS-enhanced networks escalates in environments characterized by rich scattering, such as factory settings, indoor settings, metro stations, and inside vehicles like airplanes and ships that present irregularly shaped enclosures leading to significant reverberation \cite{ref3, ref10}. Wave propagation in such environments markedly diverges from the extensively analyzed free-space scenarios. Prior research has focused on optimizing communication metrics in these settings, employing methods such as iterative experimental RIS adjustments \cite{ref10, ref13}, or leveraging available channel state information \cite{ref14}. However, the dynamic nature of these environments, driven by the movement of people and objects, introduces rapid changes and fast fading, which challenges the reliability of channel knowledge and complicates the optimization of RIS configurations\cite{ref3}. In such rich-scattering environments, wireless channels consist of a superposition of numerous reflected waves arriving from apparently random directions, rendering traditional free-space logic and methods inapplicable. The integration of sensing capabilities into RIS has emerged as a solution \cite{ref4}. It has initially been explored in simpler free-space scenarios for basic tasks like direction-of-arrival estimation for getting the channel estimate \cite{ref16}, but now increasingly required in more complex, real-world environments that demand advanced data-driven AI approaches to manage rich scattering and unpredictable propagation conditions \cite{ref8}. While leveraging RIS sensing capabilities and data-driven approaches, these approaches predominantly utilize iterative algorithms for optimizing the RIS configuration which are typically time-consuming.

In contrast to the existing works, in this paper, we introduce a robust framework designed to effectively determine the optimal RIS configuration to improve \acrshort{ue} localization accuracy within dynamically changing, rich-scattering environments using \acrfull{bilstm} \cite{ref22}. This design is adeptly suited for environments, including factories and indoor settings, where the movement trajectories of people and objects are generally predictable. Due to the formidable complexity of the rich-scattering environment, an explicit analytical approach is unfeasible. Instead, our proposed method integrates location parameters and channel sensing at RIS elements, along with channel features specific to these locations, to leverage their inherent correlations. We employ the advanced \acrshort{bilstm} model to enhance the network's ability to discern correlations across various signal contexts and establish a link between the RIS configuration and the key statistical channel parameters, i.e., channel impulse response and fading,  that significantly impacts \acrshort{ue} localization accuracy. The proposed method performs the RIS configuration with lower latency compared to the iterative algorithms.  We numerically demonstrate that the proposed BiLSTM-assisted optimization effectively tunes the RIS configuration to minimize \acrshort{mse} in \acrshort{ue} localization without needing explicit knowledge of channel statistics or their interactions with the RIS settings.


\section{System model}
\begin{figure}[t]
\centering
\centerline{\includegraphics[width=0.45\textwidth]{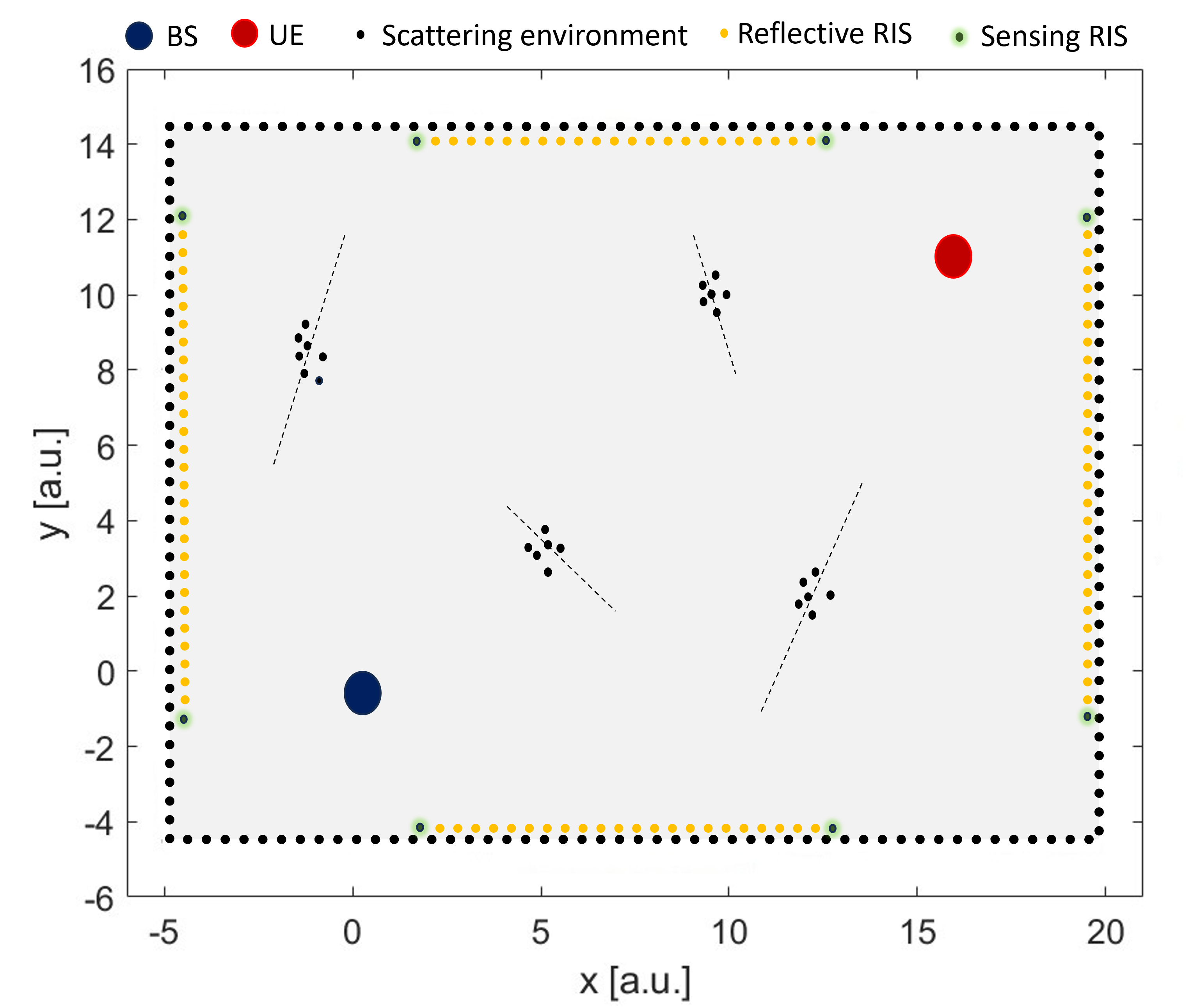}}
\caption{Considered deployment scenario in PhysFad in arbitrary units [a.u.]\cite{ref18} (Top view). An indoor scenario with a \acrshort{bs}, a \acrshort{ue}, and a distributed
RIS that partially covers the four walls of the enclosure with its eight elements dedicated to environment sensing. In addition, four \acrshort{so} moving along the indicated trajectory are also present.}
\label{fig1}
\end{figure}

We consider an enclosed indoor scenario (top view), shown in Figure \ref{fig1}, to analyse the interactions among various elements in a rich-scattering wireless environment. 
The environment causes scattering as waves reverberate and reflect multiple times off the walls. At the core of the system model lies the \acrfull{bs} that communicates at a desired frequency with the \acrshort{ue} at location \(U\). The \acrfull{so} moving along a fixed predefined trajectory are present in the environment, resulting in the dynamic nature of the fading wireless channel. The presence of \acrshort{so} introduce complexity into the system by affecting the propagation paths of the signals, resulting in a rich scattering scenario \cite{ref10}. The status of the \acrshort{so} is presented by the vector \(P\) that concatenates all essential parameters required to comprehensively describe the location status of the \acrshort{so}. The RIS operating in hybrid fashion is strategically placed on the periphery of the communication environment in a distributed manner. It consists of \(N_{\text{RIS}}\) reflective elements, each capable of adopting states from a predefined vector \(K\) of dimensions \(1 \text{x} N_{\text{RIS}}\), where \(K \in C^{N_{\text{RIS}}}\). Given that RIS systems generally have a limited set of configurations, such as those managed by PIN diodes \cite{ref2}, we assume \(C\) to be finite and binary. In addition, we assume that \(S_{\text{RIS}}\) RIS elements are dedicated to channel sensing \cite{ref3, ref4}. 

\subsection{Channel Model}
We assume that channel perturbations owing to \acrshort{so} occur rapidly enough to model the input-output relationship as a fast-fading, frequency-selective channel in discrete time. The field observed at any given point in a rich-scattering environment results from the superposition of all reflections, with reflections off specific objects varying according to their location \cite{ref4}. Consequently, a sequence of field measurements can serve as a unique wave fingerprint, facilitating the accurate determination of the \acrshort{ue} location. 

The wireless channel at the \acrshort{ue} is influenced by the RIS configuration, \(K\), and \acrshort{so} location status, \(P\). Let, \(H_{BS-UE}\) denote the channel impulse response at the \acrshort{ue}, which includes line of sight signals, and the signals that interact with the RIS, \acrshort{so} and walls of the enclosure. The influence of the RIS configuration setting \(K\) on \(H_{BS-UE}\) can be highly complex and potentially intractable \cite{ref3}. We further employ \(S_{\text{RIS}}\) RIS elements to sense and learn the channel \(H_{BS-S}\), between the \acrshort{bs} and \(S_{\text{RIS}}\), to acquire \acrshort{so} status \(P\). It is assumed that in the offline calibration phase, the system model identifies and learns the optimized RIS configurations corresponding to every potential set of \acrshort{so} locations \(P\) using the method descried in Section \ref{method}. Then, during the real-time operation, the RIS adopts the best configuration from the codebook based on the currently detected \(P\) by the \(S_{\text{RIS}}\). 

Traditional cascaded models do not adequately represent the nonlinear parametrization of the multipath wireless channel via the RIS. The training data for our \acrshort{bilstm} approach is simulated using a channel simulator, Physfad, a physics-based, end-to-end communication model tailored for RIS-parametrized wireless environments that accommodates adjustable fading and accurately reflects the wave physics of the described rich-scattering environment \cite{ref19}. Developed from first principles through a coupled-dipole formalism, PhysFad strictly adheres to fundamental physical laws by describing wireless entities as either individual dipoles or assemblies, characterized by properties such as resonance frequency and absorption. As illustrated in Figure \ref{fig1}, it models continuous surfaces of scattering objects as arrays of dipole fences, with each RIS element's configuration determined by its dipole's resonance frequency \cite{ref19}. For the sake of simplicity, we are operating in a two-dimensional space where we examine dipoles positioned within the \(x\)-\(y\) plane, with their dipole moments aligned vertically along the \(z\)-axis. PhysFad computes the wireless channel \(H_{BS-UE}\) between designated \acrshort{bs} and \acrshort{ue} dipoles, and \(H_{BS-S}\) between \acrshort{bs} and \(S_{\text{RIS}}\) dipoles, considering both RIS and \acrshort{so} in the environment. It comprehensively covers aspects such as space and causality, dispersion with frequency selectivity, and the interconnection between phase and amplitude responses of RIS elements. The model also accounts for mutual coupling effects and long-range mesoscopic correlations, alongside the nonlinear parametrization of wireless channels influenced by both RIS configurations and environmental \acrshort{so}. PhysFad is constructed using arbitrary units, with the central operating frequency, trasnmit power and the medium’s permittivity and permeability uniformly set to one, further mathematical details can be found in \cite{ref19}. 

\section{methods}
\label{method}
In this section, we formulate our problem and present the proposed solution to optimize RIS configuration, minimizing \acrshort{mse} of \acrshort{ue} localization in dynamically changing, rich-scattering environments.

\subsection{Problem Formulation}
Our objective is to develop a method for optimizing the RIS configuration, \(K\), to localize the \acrshort{ue} accurately. The localization performance is quantified by the \acrshort{mse} between the estimated \((\hat{x},\hat{y})\)and actual \((x,y)\) locations, i.e., ground truth of the UE. This is mathematically expressed as

\begin{equation}
\text{MSE} = \frac{1}{N} \sum_{i=1}^{N} \left(Y_i -  f(Y_i) \right)^2= \frac{1}{N} \sum_{i=1}^{N}\left( (\hat{x}_i - x_i)^2 + (\hat{y}_i - y_i)^2 \right)
\label{eq1}
\end{equation}
where \(N\) is the number of measurement points, \( f(Y_i) \) denotes the estimated coordinates, \( Y_i \) represents the true coordinates, \((\hat{x}_i, \hat{y}_i)\) represent the estimated coordinates of the UE at the \(i\)-th measurement, and \((x_i, y_i)\) are the actual coordinates. As the channel experiences fast-fading, our focus is on minimizing the \acrshort{mse} in UE localization, which involves adjusting \(K\) based on understanding of the channel's stochastic behavior influenced by this configuration. 

In free-space operations, the wireless channel \(H_{BS-UE}\) is typically divided into a straightforward linear cascade: from the \acrshort{bs} to the RIS, through the RIS configuration, and from the RIS to the \acrshort{ue}. However, this model falls short in rich-scattering environments where each signal path might interact with various RIS elements along its complex trajectory. Consequently, the parametrization of the channel \(H_{BS-UE}\) involving RIS and dynamic \acrshort{so} becomes highly complex and analytically intractable. To tackle this challenge, we use a learned \acrshort{bilstm}. Since the direct correlation between the UE location \(U\), \(H_{BS-UE}\) , \(H_{BS-S}\), and \(K\) is unknown, highly intricate, and influenced by the status of \acrshort{so}, \(P\), we developed a training dataset comprising \(H_{BS-UE}\), \(H_{BS-S}\), \(P\) and, respective configurations \(K\) for each user location \(U\). This dataset includes \(L\) pairs in the format \(\{(H_{BS-UE})_l, (H_{BS-S})_l, P_l, K_l, U_l\}_{l=1}^L\). This dataset enables us to precisely fine-tune the RIS configuration, thereby achieving optimal UE localization with minimal \acrshort{mse}. Through intelligent adaptation of the RIS settings in response to observed channel variations, we systematically minimize the MSE, enhancing localization accuracy.

\subsection{RIS Configuration Optimization}
\label{config}
Our proposed approach for optimizing the RIS configuration \(K\), based on the UE localization objective (\ref{eq1}), employs a BiLSTM model. This is achieved by training the BiLSTM to recognize patterns from historical data \(\{(H_{BS-UE})_l, (H_{BS-S})_l, P_l, K_l, U_l\}_{l=1}^L\),  learning the relationships between input features, \(\{H_{BS-UE}, H_{BS-S}, P\}\) and their corresponding outcomes \(\{ K, U\}\). 

In order to learn how to optimize \(K\) with the estimate of \(U\), we use a \acrshort{bilstm} model, termed as \(\theta\), whose exact architecture employed in the experimentation is explained in detail in Section \ref{results}. The model is designed to generate two combined outcomes: it predicts the \acrshort{ue} location coordinates as a regression problem as well as the RIS configuration as a classification problem that results in the minimum of (\ref{eq1}). The model employs a hybrid loss function to cater to the dual objectives of the network. For the regression output, i.e., predicted UE coordinates, \(\hat{U}\), we use the \acrshort{mse}, and for the classification output, \(\hat{K}\), we use Categorical Crossentropy, such that the combined loss function is given as
\begin{equation}
L(\theta) = \frac{1}{N} \sum_{i=1}^N \left(Y_i -  f_\theta(Y_i) \right)^2 + \sum_{i=1}^N \sum_{m=1}^M (-y_m \log(p_m)) + \alpha \|\theta\|^2
\end{equation}
here, \( f_\theta(Y_i) \) denotes the predicted coordinates by the \acrshort{bilstm} parameters \(\theta\), \( Y_i \) represents the true coordinates, and \(N\) is the number of data points. \( y_m \) are the true class labels, \( p_m \) are the predicted probabilities for each class, and \(M\) is the number of classes. We have added \( \alpha \) as a $l_2$-regularization parameter, which is used to avoid overfitting.

\subsection{Discussion}
The proposed closed-loop approach, where the RIS configuration is adjusted based on environmental changes, demonstrates significant advantages in scenarios where the \acrshort{so} parameter space is manageable, streamlining real-time operational decisions by quickly referencing pre-stored configurations from a codebook. This method bypasses the need for learning complex nonlinear channel parametrizations directly, instead utilizing a codebook that matches observed environmental conditions with optimal configurations, thus minimizing latency and computational overhead. However, it requires precise initial calibration and faces challenges due to the necessary discretization of continuous \acrshort{so} parameters, i.e., location, which can lead to performance losses if the codebook resolution is not sufficiently fine. The degree to which a \acrshort{so} affects the wireless channel further affects the accuracy of the \acrshort{bilstm} to estimate \(U\), even with limited measurement sequences or in noisy conditions \cite{ref20}. BiLSTM networks are particularly suited for this task over other neural network architectures due to their ability to process sequence data in both forward and backward directions, providing a richer understanding of context and temporal relationships within the data \cite{ref22}. This capability is critical in dynamic environments where the influence of past and future data points can provide crucial insights for accurate prediction and optimization. 

\section{Results}
\label{results}
We have implemented our proposed method in a numerical study where a rich-scattering environment, shown in Figure \ref{fig1}, was simulated using \cite{ref19}. This setup includes a single-antenna \acrshort{bs}, a single-antenna \acrshort{ue} with changing location, and a RIS equipped with \(N_{\text{RIS}} = (20, 60, 100) \), binary-tunable elements. For clarity, the specific parameters of the dipoles, the resonance frequency (\(f_{\text{res}}\)), charge term (\(\chi\)), and absorptive damping term (\(\Gamma_L\)), utilized in our PhysFad simulations are set for transceivers as \(\{f_{\text{res}}=1, \chi=0.5, \Gamma_L=0\}\). In the scattering environment, these parameters are \(\{f_{\text{res}}=10, \chi=50, \Gamma_L=10^4\}\) with dipole spacing of one-fourth of the wavelength, $\lambda$. Lastly, for the RIS, the parameters are \(\{f_{\text{res}}=\{1, 5\}, \chi=0.2, \Gamma_L=0.03\}\).

The data preparation phase includes acquiring frequency response measurements and \acrshort{so} status across various RIS profiles and \acrshort{ue} locations, as detailed in Section \ref{config}. For a fixed UE location and a fixed \(N_{\text{RIS}}\), different  RIS configuration were considered. For each configuration, the \acrshort{so} was located at 10 randomly chosen locations along their fixed trajectory of motion, from which channel response measurements \(H_{BS-UE}\) and \(H_{BS-S}\)  were obtained. RIS configurations \(K_l\) are one-hot encoded to facilitate classification tasks within the network architecture. The dataset, represented as \(\{(H_{BS-UE})_l, (H_{BS-S})_l, P_l, K_l, U_l\}_{l=1}^L\), was divided into 80\% for training and 20\% for testing. For validation purposes the training set is further divided into 80\% training and 20\% validation which assists in hyper parameter optimization using grid search to fine tune the model settings and helps prevent overfitting.

The designed \acrshort{bilstm} incorporates a dual-input architecture. It employs a \acrshort{bilstm} layer with 50 units to capture temporal dependencies from sequential channel data, followed by an additional \acrfull{lstm} layer with 50 units using ReLU activation to further refine the features. A separate input for RIS configuration features includes an embedding layer that transforms categorical inputs into a 20-dimensional vector, subsequently flattened for processing. The outputs from both \acrshort{lstm} and embedding layers are concatenated and directed into two dense output layers—one with linear activation for predicting \acrshort{ue} coordinates and another with softmax activation for classifying RIS configurations. The model was compiled using the Adam optimizer \cite{ref21} with a learning rate of 0.0001, applying \acrshort{mse} as the loss function for coordinates and categorical cross-entropy for RIS configuration classification. Training spanned 200 epochs with a batch size of 32, and included a checkpointing mechanism to save only the model iteration with the lowest validation loss, ensuring the highest performance during final evaluations on the test set. Furthermore, for the sake of comparison and to establish a baseline \acrshort{mse}, a simple neural network was trained with random RIS configuration to predict user location given the same dataset splits.


\begin{figure}[t]

\centering
\centerline{\includegraphics[width=0.46\textwidth]{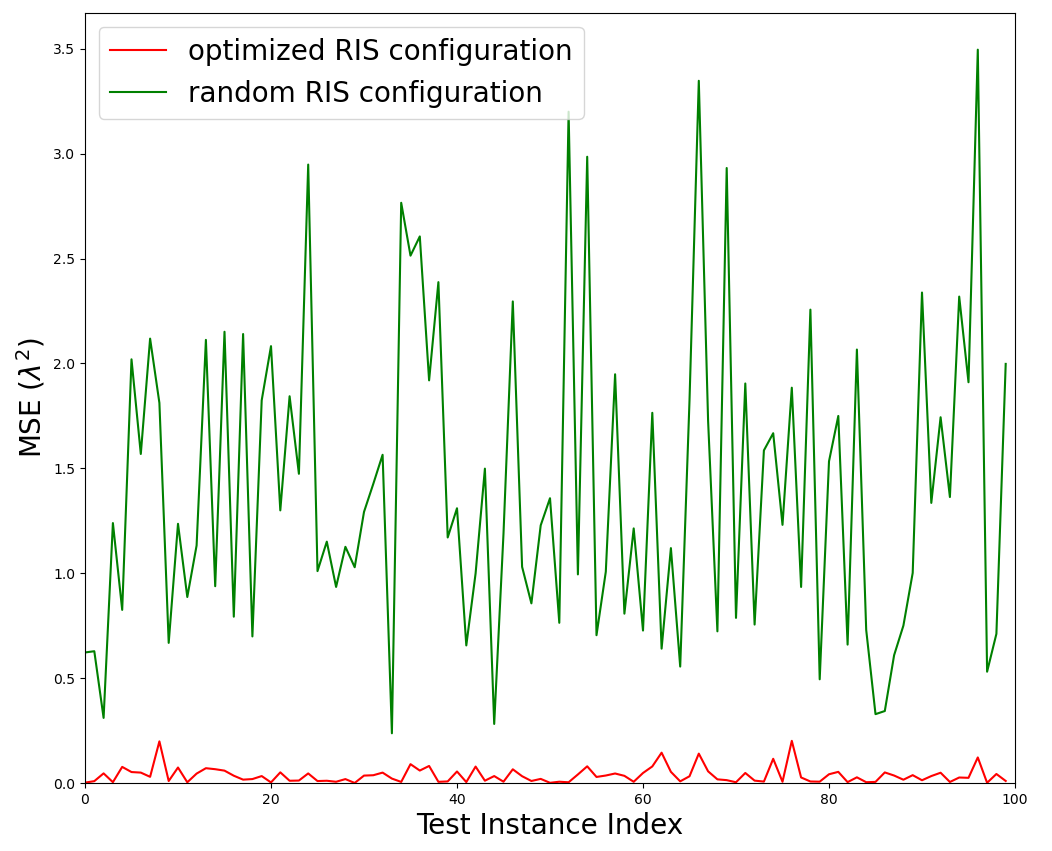}}
\caption{MSE versus test index for random and optimized RIS configuration with \(K=500\) and \(N_{\text{RIS}}=20\).}
\label{fig2}
\end{figure}
\begin{figure}[t]
\centering
\centerline{\includegraphics[width=0.46\textwidth]{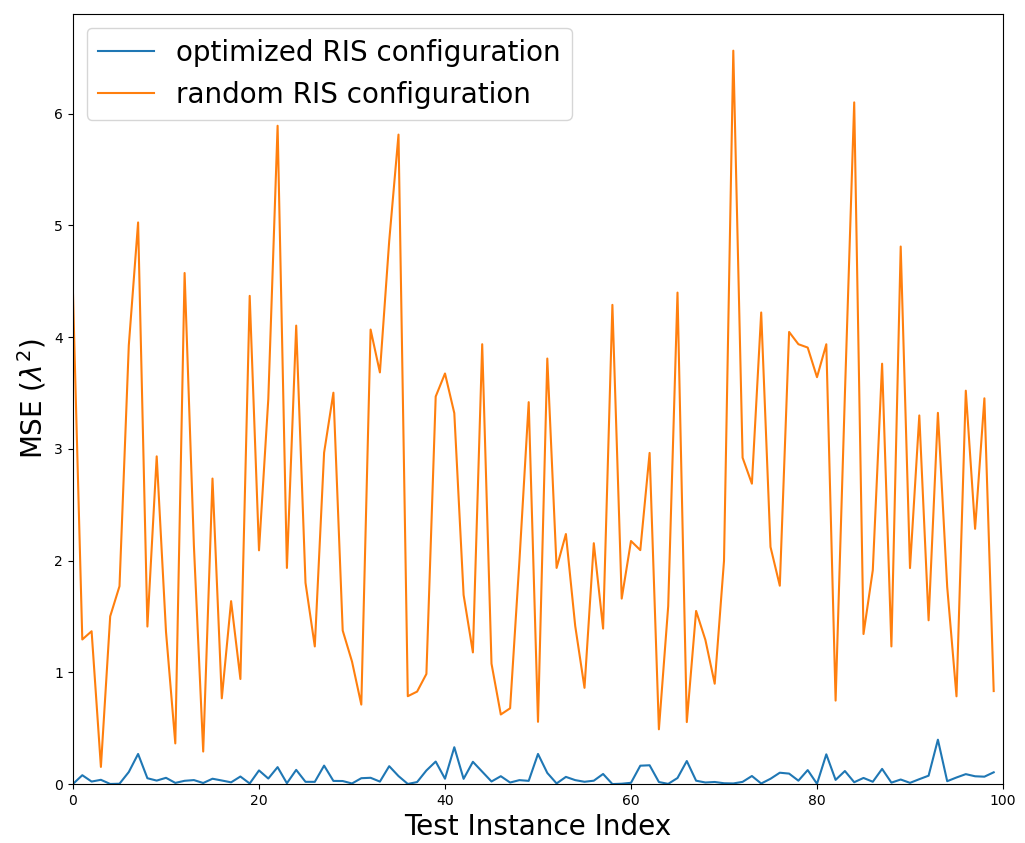}}
\caption{MSE versus test index with optimized and random RIS configuration with \(N_{\text{RIS}}=100\) and \(K=100\). }
\label{fig3}
\vspace{-2mm}
\end{figure}

After extensive training of the \acrshort{bilstm} model, we evaluated its efficacy in optimizing RIS configurations for UE localization, aiming to minimize the \acrshort{mse}. This evaluation was conducted using 100 randomly selected test instances. We employed \(N_{\text{RIS}} = 20\) and three distinct RIS configurations, \(K=\{10,100,500\}\), to assess the model's performance. The results, depicted in Figure \ref{fig2}, exhibit a high degree of accuracy in determining the UE location with proposed RIS configuration optimization method compared to the random RIS configuration. Results for varying $K$ are summarized in Table \ref{table1}, where we see that increasing \(K\) from 10 to 500 improved the performance by approximately 34.39\%. Notably, the accuracy of the localization will improve significantly as the model is trained on data rich in RIS configurations, i.e., in our case \(2^{N_{\text{RIS}}}\) distint configuration are possible. 

\begin{table}[t]
\caption{Average MSE for random and optimized RIS configuration, standard deviation $\sigma$ of optimized MSE, and \% error reduction compared to baseline.}
\centering

\begin{tabular}{|c|c|p{1.2cm}|p{1.2cm}|c|p{1.5cm}|}
\hline
\textbf{\(N_{\text{RIS}}\)} & \textbf{K} & \textbf{Baseline MSE ($\lambda^2$)} & \textbf{Optimized \acrshort{mse} ($\lambda^2$)} & \textbf{$\sigma$} & \textbf{\% error reduction} \\
\hline
20 & 10 & 0.94 & \textbf{0.35} & 0.271 & 62.77\% \\
\hline
20 & 100 & 0.27 & \textbf{0.08} & 0.086 & 70.37\% \\
\hline
20 & 500 & 1.41 & \textbf{0.04} & 0.038 & \textbf{97.16}\% \\
\hline
60 & 100 & 0.59 & \textbf{0.17} & 0.166 & 71.19\% \\
\hline
100 & 100 & 2.45 & \textbf{0.07} & 0.075 & \textbf{97.14}\% \\
\hline
\end{tabular}
\vspace{-2mm}
\label{table1}
\vspace{-2mm}
\end{table}
We further investigated the influence of the varying number of \acrshort{ris} elements, \(N_{\text{RIS}}=\{20,60,100\}\), on the localization accuracy of system model using proposed \acrshort{bilstm} network, as illustrated in Figure~\ref{fig3} and summarized in Table \ref{table1}. The analysis shows that while the \acrshort{mse} decreases as \(N_{\text{RIS}}\) increases, the improvement in \acrshort{mse} is relatively modest. This could be attributed to the same number of configurations used in the model, regardless of the significant increase in \(N_{\text{RIS}}\), despite the exponential growth in the potential number of configurations, theoretically \(2^{N_{\text{RIS}}}\). It suggests that the full potential of increasing \(N_{\text{RIS}}\) is not being harnessed. Further enhancement in \acrshort{mse} could likely be achieved by utilizing a broader variety of \acrshort{ris} configurations, particularly in scenarios involving frequent changes in the location of the \acrshort{so}. Hence, optimizing the selection and deployment of \(N_{\text{RIS}}\) is critical, and should consider the environmental dynamics, including the size, number, and mobility of \acrshort{so}, to maximize the system performance. 

\section{Conclusion}
In this paper, we address the challenges posed by dynamically changing rich-scattering environments with a novel RIS configuration algorithm that forgoes traditional analytical modeling in favor of learning directly from data. Our method utilizes a \acrshort{bilstm} model to dynamically optimize RIS configurations, for enhancing UE localization accuracy by minimizing the \acrshort{mse}. Numerical evaluations demonstrate the effectiveness of this approach, showcasing its capability to fine-tune RIS settings to improve localization accuracy in complex scattering conditions. Proposed approach holds potential for jointly optimizing communication parameters within a unified framework encompassing communication, sensing, and localization tasks. The exploration of this integrated approach has been left  for future work.

\section*{Acknowledgment}
This project has received funding from the European Union’s Horizon Europe Research Program under grant agreement No. 101058505 - 5G-TIMBER, and from the Estonian Education and Youth Board ÕÜF11 \"AIoT*5G - Artificial intelligence, edge computing and IoT solutions in distributed systems.

\bibliographystyle{ieeetr} 
\bibliography{references}
\end{document}